# A Multilayer Neural Network Merging Image Preprocessing and Pattern Recognition by Integrating Diffusion and Drift Memristors

Zhiri Tang, Ruohua Zhu, Ruihan Hu, Yanhua Chen, Edmond Q. Wu, Hao Wang, Jin He, *Senior Member, IEEE*, Qijun Huang, and Sheng Chang, *Senior Member, IEEE*

*Abstract*—With the development of research on novel memristor model and device, neural networks by integrating various memristor models have become a hot research topic recently. However, state-of-the-art works still build such neural networks using drift memristor only. Furthermore, some other related works are only applied to a few individual applications including pattern recognition and edge detection. In this paper, a novel kind of multilayer neural network is proposed, in which diffusion and drift memristor models are applied to construct a system merging image preprocessing and pattern recognition. Specifically, the entire network consists of two diffusion memristive cellular layers for image preprocessing and one drift memristive feedforward layer for pattern recognition. Experimental results show that good recognition accuracy of noisy MNIST is obtained due to the fusion of image preprocessing and pattern recognition. Moreover, owing to high-efficiency in-memory computing and brief spiking encoding methods, high processing speed, high throughput, and few hardware resources of the entire network are achieved.

This work was supported by the National Natural Science Foundation of China (61874079, 61574102, 61671293 and U1933125), the Fundamental Research Fund for the Central Universities, Wuhan University (2042017gf0052), the Wuhan Research Program of Application Foundation and Frontier Technology (2018010401011289), and the Luojia Young Scholars Program. Part of calculation in this paper has been done on the supercomputing system in the Supercomputing Center of Wuhan University. (*Corresponding author: Sheng Chang*)

Zhiri Tang is with the School of Physics and Technology, Wuhan University, Wuhan, China. He is also with the Department of Computer Science, City University of Hong Kong, Hong Kong, China. (E-mail: gerin.tang@my.cityu.edu.hk).

Ruohua Zhu is with School of Physics and Electronics, Henan University, Kaifeng, China.

Ruihan Hu is with Guangdong Key Laboratory of Modern Control Technology, Guangdong Insitute of Intelligent Manufacturing, Guangzhou, China.

Yanhua Chen is with Department of Geography, The University of Hong Kong, Hong Kong, China.

Edmond Q. Wu is with Department of Automation, Shanghai Jiao Tong University, Shanghai, China.

Hao Wang, Jin He, Qijun Huang, and Sheng Chang are with the School of Physics and Technology, Wuhan University, Wuhan, China. (E-mail: changsheng@whu.edu.cn).

## I. INTRODUCTION

L. O. Chua postulated memristor [1] in 1971, which is the fourth basic element besides resistor, inductor, and capacitor. In general, memristor can be divided into two main types [2]: drift memristor and diffusion memristor. HP Lab made drift memristor in 2008 first [3] while diffusion memristor was first presented by UMass in 2016 [4]. Since memristor device has been realized, research related to memristor have emerged one after another. Recently, neuromorphic computing with memristive neural networks (MNNs) [5]-[6], which uses memristor as artificial synapse [7], has become a hot research topic due to memristor's potential on many combinatorial descriptions of biological synapse's characteristics [8], such as spike timing-dependent plasticity (STDP) [9], long-term potentiation (LTP) [10], and long-term depression (LTD) [11]. Furthermore, memristor has many advantages, such as in-memory computing [12], high integration density [13]-[14] ,and small dimension [15]. Hence, memristive neural networks show great potentials on various learning frameworks including long short-term memory network (LSTM) [16], reinforcement learning (RL) [17], and recurrent convolutional networks [18], and many applications including pattern recognition [19], edge detection [20], temporal data classification [21], and high performance computing [22]-[23].

Although the number of existing works about memristor are growing, most research and applications still used the drift memristor only [24]-[25]. In these years, the emergence of diffusion memristor gives room for the development of MNNs with richer functions and stronger performance. Q. F. Xia and J. J. Yang described diffusion memristor as a synaptic emulator for neuromorphic computing [4] when they first time made the diffusion memristor device in UMass. Z. Wang applied diffusion memristor into artificial nociceptor [26] and pattern recognition [27], in which diffusion memristors were used as threshold switches and crossbar structure, respectively. However, how to explore the potentials of diffusion memristor in other fields is still a challenging problem.

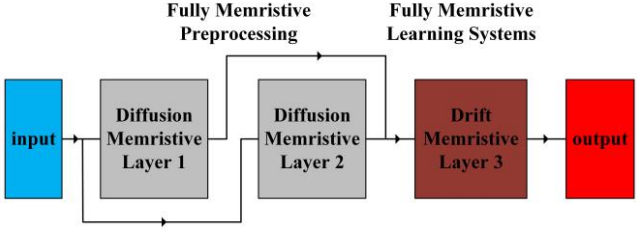

Fig. 1. Overview of entire network

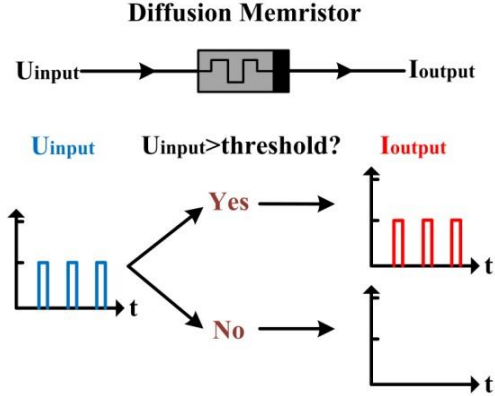

Fig. 2. Switching characteristic of diffusion memristor model

According to state-of-the-art works, how to design a neural network by integrating drift and diffusion memristor models is also an open area. One research about fully memristive neural networks used memristors and metal-oxide-semiconductor field effect transistors (MOSFET) [28] while another built a fully memristive crossbar by memristors, capacitors, and inductors [27]. From the above research, other devices besides memristor are still needed in fully memristive neural networks, which limit not only improvements on scalability [29] and processing speed [30], but also simplifications on fabrication process [31] of the entire networks.

Furthermore, existing research about MNNs in image processing have many different applications, such as edge detection [32] and image storage [33]. However, there are no research on how to combine image preprocessing with learning systems and explore the potentials of diffusion memristor in image processing. Theoretically, image preprocessing with MNNs can improve the performance of entire systems including recognition accuracy, processing speed, and hardware resource. Hence, a novel neural network framework, which use drift and diffusion memristor models only, and combine image preprocessing with learning systems smoothly, need to be developed.

Inspired by the above, this paper designs a multilayer neural network merging image preprocessing and learning systems

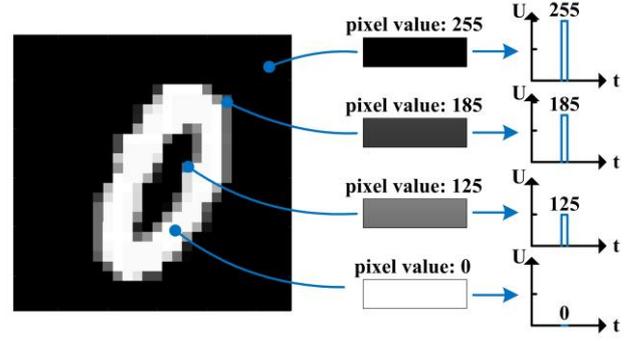

Fig. 3. The way pixels are converted into spikes

with diffusion and drift memristor models only. The main contributions are summarized as follows:

a. A novel multilayer neural network is proposed, which merges image preprocessing with learning systems. Through this way, a fully memristive system with high-efficiency in-memory computing and brief spiking encoding methods is implemented, which lays a solid foundation for achieving a good performance on machine learning tasks.

b. To verify its merits, the proposed framework is applied to noisy pattern recognition on hardware. Image denoising and pattern recognition are implemented by image preprocessing and learning systems, respectively. Experimental results show that it achieves high noisy pattern recognition accuracy with high processing speed, high throughput, and few hardware resources.

## II. OVERVIEWS OF ALGORITHMS

In this section, a novel multilayer neural network with function of image preprocessing and learning systems is proposed, which consists of two diffusion memristive layers for image preprocessing and one drift memristive layer for pattern recognition shown as Fig. 1. To verify the function of image preprocessing and learning systems, the entire network is applied to denoising and pattern recognition for noisy image.

### A. Diffusion Memristive Cellular Layer for Image Preprocessing

The basic diffusion memristor model was proposed by researchers from UMass [4] and has similar characteristics to a threshold switch, which is shown as Fig. 2.

If the hold time of input voltage spikes of diffusion memristor model is the same and the spikes exceed the set threshold, the diffusion memristor model will "open" and the output current spikes will follow the input voltage spikes. Due to its switching characteristics, diffusion memristor model can be used as a natural mean filter. The basic diffusion memristor model with the same hold time of input voltage spikes is as follow:

$$\begin{cases} I_{output}(t) = U_{input}, & U_{input} > Threshold \\ I_{output}(t) = 0, & U_{input} \leq Threshold \end{cases} \quad (1)$$

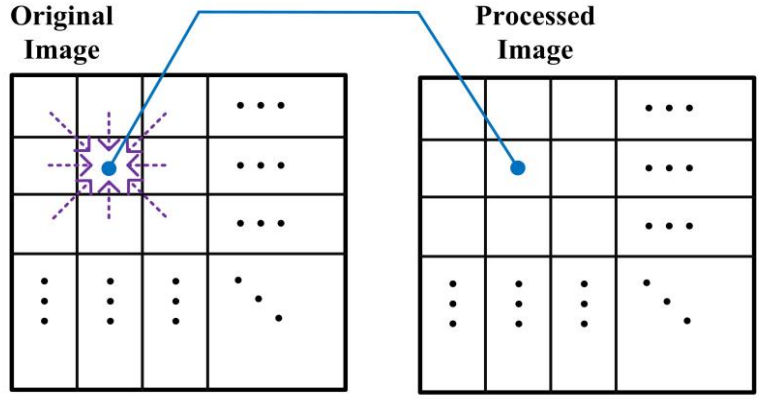

Fig. 4. The image preprocessing of diffusion memristive cellular layer

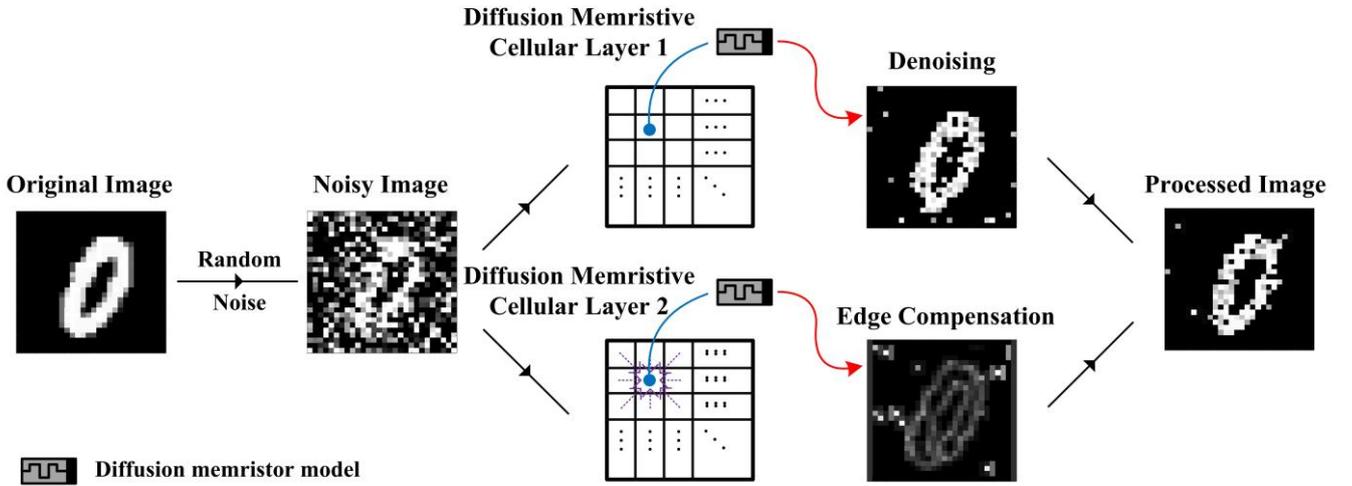

Fig. 5. The training process of diffusion memristive cellular layer 1 and 2 for image preprocessing

To introduce the spiking encoding methods, a black and white binary image is taken as an example, which is shown as Fig. 3. First, pixels in the image need to be converted into spikes and the hold time of each spike keeps same. Then, the pixel value corresponds to the height of its peak voltage spike. If the pixel is black, whose pixel value is 255, the spike height of this pixel will be 255. If the pixel is white, the height will be 0.

The diffusion memristive cellular layers, which has the functions of denoising and edge compensation for image preprocessing, use the eight pixels around the target pixel for maintaining high processing speed as Fig. 4.

The entire preprocessing part is shown as Fig. 5. It can be divided into two parts: memristive cellular layer 1 for denoising and layer 2 for edge compensation. As a natural mean filter, diffusion memristor models will filter out the edge information of original image when they filter most random noise. Some works [20],[32] have proposed that memristive cellular neural networks have a good performance in edge detection. Hence, layer 2 is designed for supporting the function of edge compensation through adding the edge detection results of original image to the denoising image after layer 1. In this way, while getting a good denoising performance, the original image information is also restored as much as possible.

The diffusion memristive cellular layer 1 is act as a mean filter for image denoising. The filter threshold is:

$$Threshold_{layer1} = \frac{1}{8} \sum_{f \in S} f(i,j) \quad (2)$$

The fire thresholds of each diffusion in layer 1 are set according to formula (2) and the output from layer 1 follows the formula (1).

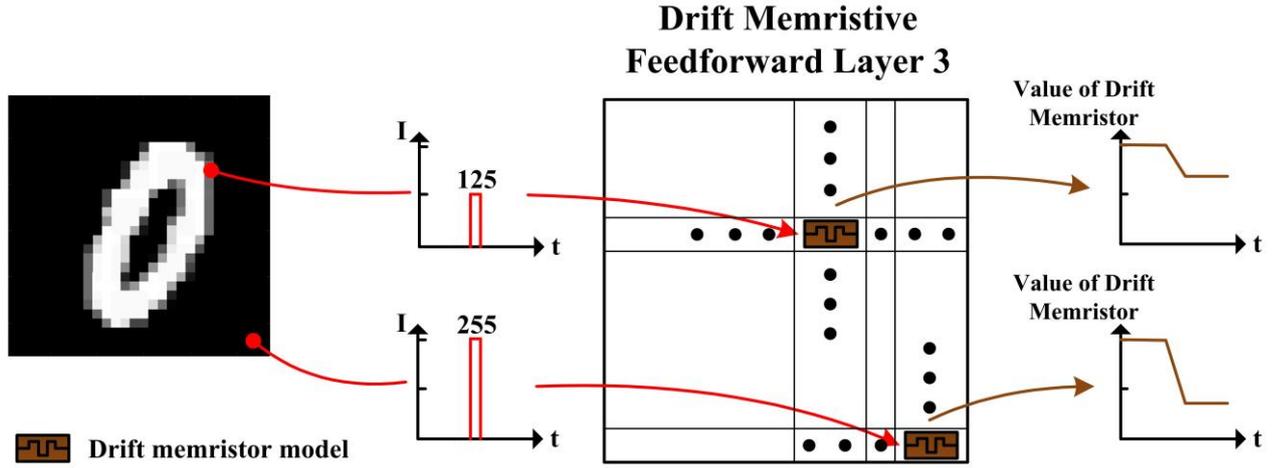

Fig. 6. The training process of drift memristive feedforward layer 3 for pattern recognition

The edge compensation of diffusion memristive cellular layer 2 follows the fully memristive cellular neural networks in [32] and the threshold for gray images is calculated by:

$$Threshold_{layer2} = R \cdot 0.299 + G \cdot 0.587 + B \cdot 0.114 \quad (3)$$

in which R, G, and B are the minimum changes of red, green, and blue that human eyes can recognize to color images, respectively.

The fire thresholds of each diffusion memristor model in layer 2 have been set separately according to formula (3). Then it will compare the surrounding eight pixels with the target pixel in turn. The voltage spikes of surrounding pixel and target pixel are applied to two ends of diffusion memristor model, respectively. Hence, if the difference between these two voltage spikes of pixels is greater than threshold, this diffusion memristor model will output a current spike. If the difference is less than threshold, this diffusion memristor model won't fire according to formula (1). After eight comparisons, the sum of output current is the processed pixel through this cellular layer. The process ends after each pixel of the image has been compared with the surrounding eight pixels.

To test the preprocessing performance of diffusion memristive cellular layer, random Gaussian noise is added to original images. The denoising preprocessing system uses one image of number 0 from MNIST and parameters of Gaussian noise are:

$$Noise = N(0, 10^4) \quad (4)$$

which has $\mu = 0$ and $\sigma = 100$.

From Fig. 5, the processed image shows a good denoising and restoration performance through adding the denoising from layer 1 with the edge compensation from layer 2. Hence, the preprocessing layers also have the same denoising effect for images of other numbers in MNIST.

*B. Drift Memristive Feedforward Layer for Pattern Recognition*

In HP memristor model, the value of drift memristor $M(t)$ can be calculated as:

$$M(t) = \frac{d(\phi(t))}{d(q(t))} = \frac{U(t)}{I(t)} \quad (5)$$

The relationship between the voltage at two ends of memristor drift model $U(t)$ and the current through drift memristor model $I(t)$ is:

$$U(t) = I(t)[R_{off} - (R_{off} - R_{on})\mu_v \frac{R_{on}}{D^2} \int_{-\infty}^{t} I(t)dt] \quad (6)$$

So we have:

$$M(t) = R_{off} - (R_{off} - R_{on})\mu_v \frac{R_{on}}{D^2} \int_{-\infty}^{t} I(t)dt$$

$$= k_1 - k_2 \int_{-\infty}^{t} I(t)dt \quad (7)$$

where $k_1 = R_{off}$ and $k_2 = (R_{off} - R_{on})\mu_v \frac{R_{on}}{D^2}$.

The change of synaptic weight equals to change of reciprocal value of drift memristor model [4, 7, 11], which has:

$$dW = dG = d(\frac{1}{M(t)}) = d(\frac{1}{k_1 - k_2 \int_{-\infty}^{t} I(t)dt}) \quad (8)$$

Due to the way spikes encoded, $I(t)$ denotes the constant for a specific pixel and $I(t)dt$ is the electricity flowing through this drift memristor model, which is the area of spikes in the coordinate system. So we have:

$$M(t) = k_1 - k_2 \cdot Area_{spike} \quad (9)$$

$$dW = dG = d(\frac{1}{M(t)}) = d(\frac{1}{k_1 - k_2 Area_{spike}}) \quad (10)$$

Further, considering non-idealities of the memristors, a complex memristor model which has window function is shown as follow:

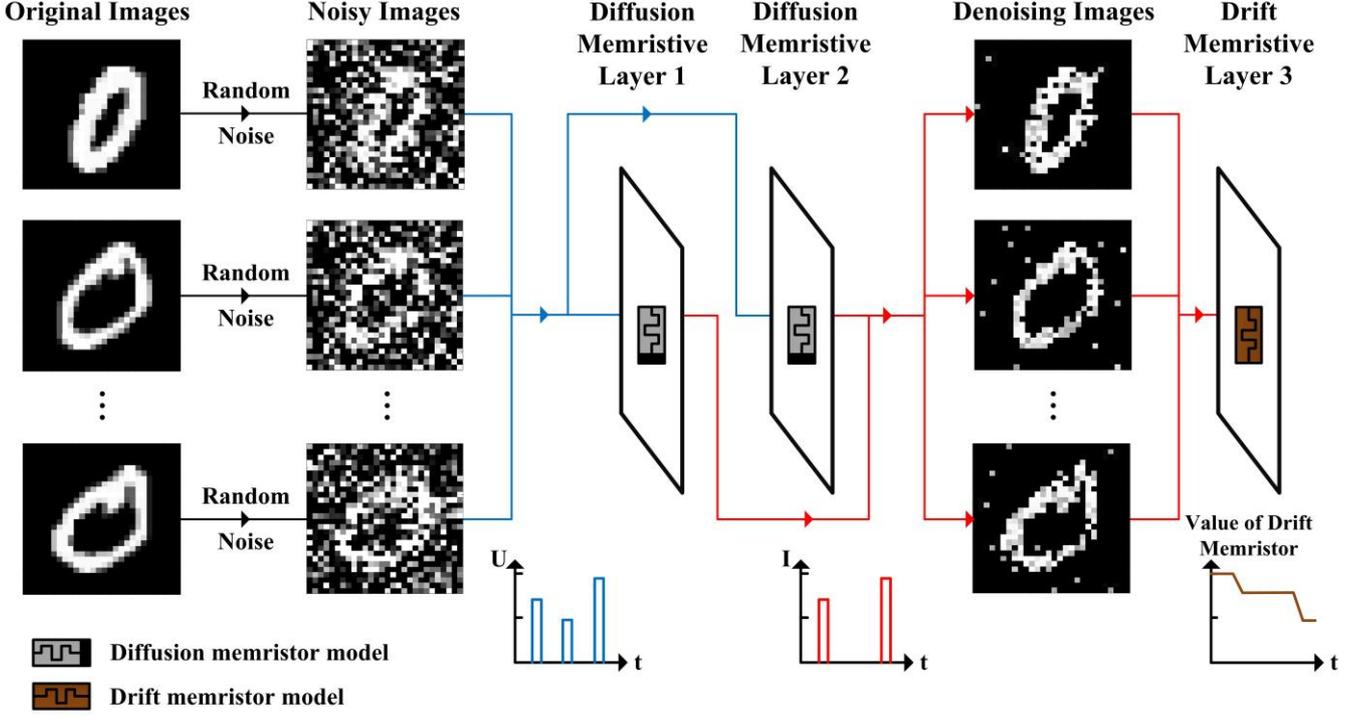

Fig. 7. The architecture and training process of entire network

$$W(t) = \int_{-\infty}^{t} \mu_v \frac{R_{on}}{D} I(t) f(x) dt = \mu_v \frac{R_{on}}{D} q(t) f(x)$$
(11)

So the value of drift memristor model can be derived as:

$$M(t) = R_{on}\left(\mu_v \frac{R_{on}}{D^2} f(x) \int_{-\infty}^{t} I(t) dt\right)$$
$$+ R_{off}\left(1 - \mu_v \frac{R_{on}}{D^2} f(x) \int_{-\infty}^{t} I(t) dt\right)$$
$$= R_{off} - (R_{off} - R_{on})\mu_v \frac{R_{on}}{D^2} f(x) \int_{-\infty}^{t} I(t) dt$$
(12)

From above, the drift memristor model with non-idealities can also be transformed into formula (9). Hence, our network has nothing to do with the non-idealities of memristor models. In other words, the framework is universal to all kinds of complex or nonlinear memristor models because all our network needs is just the basic characteristic of memristor models.

From above, the value of drift memristor model keeps a linear relationship with the total area of spikes. If input spikes of one drift memristor model are from black pixels, the drift memristor model corresponding to these black pixels will be lower than that corresponding to white pixels, which is a brief and efficient training process for one drift memristor model.

The total drift memristive feedforward layer is shown as Fig. 6. Each pixel has one drift memristor model to act as its synaptic weight. If one image is input into the feedforward layer during the training process, all the values of drift memristor models will be changed.

During the inference process, the pixels of inference images are converted into voltages whose value depends on its pixel. The "voltage" image is applied to each drift memristor model in layer 3. Due to the difference in drift memristor models after training, the output currents of each pixel will be different, and the sum of output currents will be used as numerical answers for inference.

*C. Entire Network Architecture*

The architecture and training process of entire network are shown as Fig. 7. First, random Gaussian noises are added to original images from MNIST. Then noisy images are converted into voltage spikes, which are represented by blue lines. Second, these voltage spikes are input into diffusion memristive layer 1 and layer 2, respectively. The output current spikes from layer 1 and layer 2, which are represented by red lines, are input into drift memristive layer 3 together. Finally, the values of drift memristor models in layer 3 will be changed during the training process.

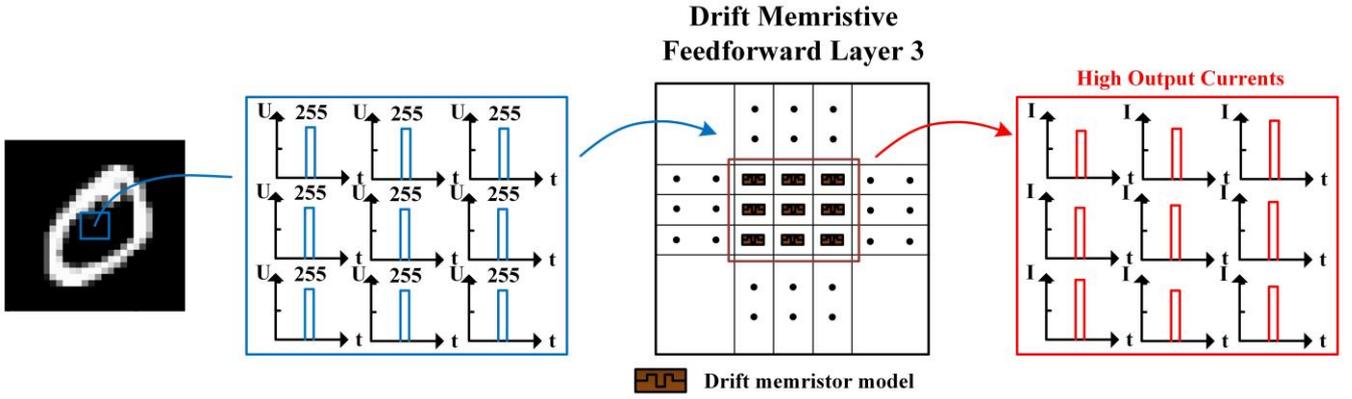

Fig. 8. The inference process of entire network

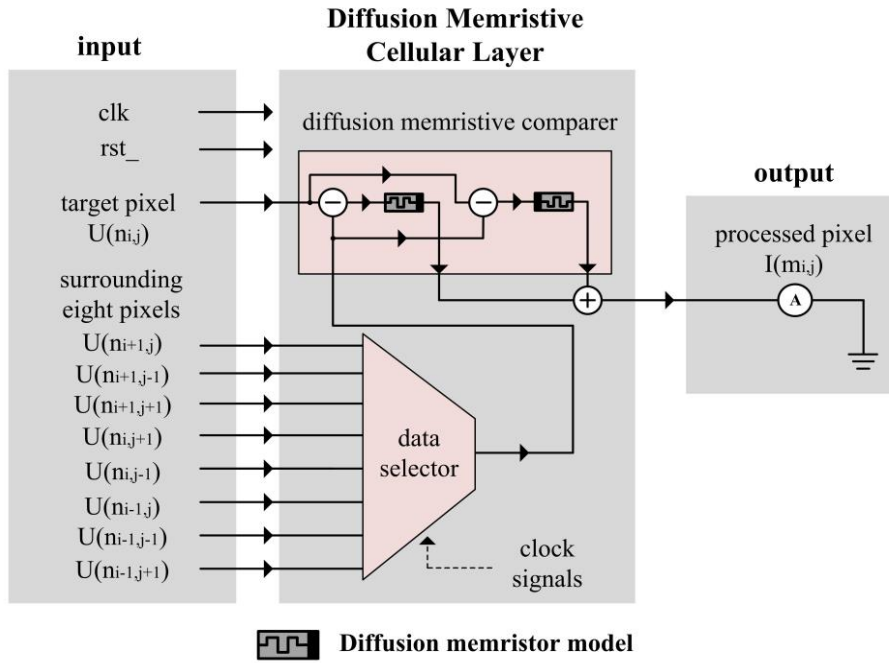

Fig. 9. The circuit diagram of diffusion memristive cellular layer for image preprocessing

The inference process of entire network needs drift memristive layer 3 only. From above, the values of drift memristor models vary after training. Take an area of random black pixels with high voltages as an example, the output of these random pixels will be higher because the values of drift memristor models in this area are lower than others after training, which is shown as Fig. 8. For a wrong test image, the input in the area contains some pixels being or closed to white, which will result in lower output currents. Hence, the sum of output currents corresponding to right category is larger than other wrong categories. In other words, the right test images have a better "match" for the drift memristive layer 3.

### III. OVERVIEWS OF HARDWARE DESIGN

In this section, circuit diagrams of diffusion memristive layer, drift memristive layer, and entire network are introduced based on diffusion and drift memristor models on Field Programmable Gate Array (FPGA). Some hardware optimization techniques are used to improve the hardware performance including processing speed, throughput, and hardware resource. Further, to better show the circuit framework of the entire system, we use a single memristor model to represent the corresponding memristive layer.

*A. Circuit Diagram of Diffusion Memristive Cellular Layer*

To build the entire network and verify the network's performance on hardware, FPGA is chosen as hardware platform. FPGA is controlled by clock signals so that it is very suitable to process spiking signals in many applications [34-37]. First, the diffusion memristor is modelled on FPGA according to formula (1). Based on the model, the circuit diagram of diffusion memristive cellular layer for image preprocessing is shown as Fig. 9. The input includes clock signals, reset signals, and voltages of target pixel and eight surrounding pixels, in which the input voltage of pixels depend on the value of corresponding pixels, e.g. input voltage is 2.5mV for whose pixel value is 25 and 14.5mv for whose pixel value is 145. The output is the processed pixel composed by current spikes, which ranges from 0 to 229.5mA according to formula (1) to (3) and the framework of diffusion memristive cellular layers. In diffusion memristor model comparer, each pixel has two diffusion memristor models for positive and negative comparisons. Clock signal-controlled data selector determines the order of comparisons. The entire memristive cellular layer is pipelined without any complex calculation inside. Hence, the processing speed of the entire system is quite high theoretically.

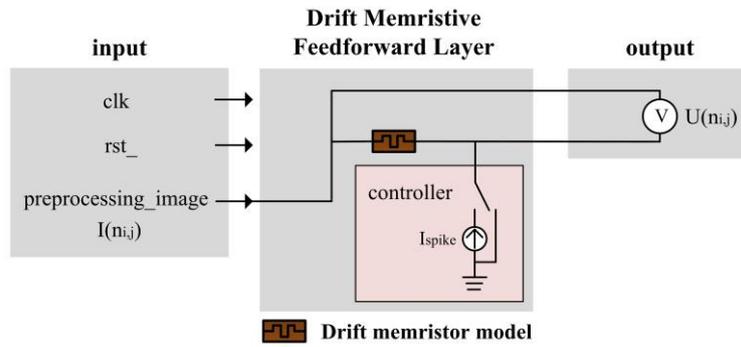

Fig. 10. The circuit diagram of drift memristive feedforward layer for pattern recognition

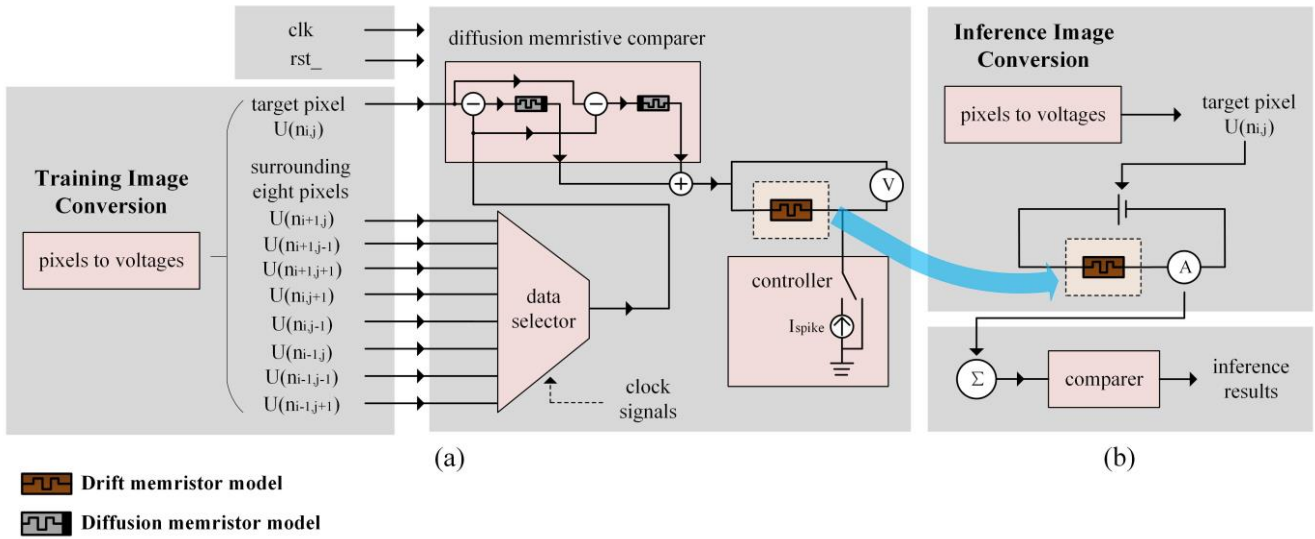

Fig. 11. The circuit diagram of entire network's training and inference

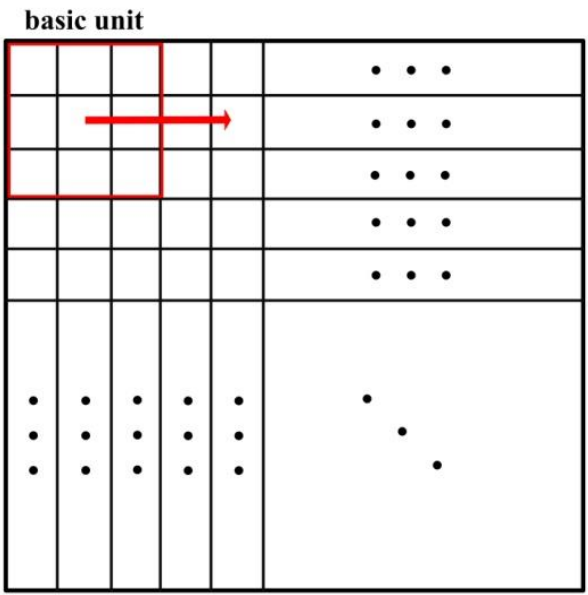

Fig. 12. Using basic unit of nine pixels to perform loading and calculation operations

*B. Circuit Diagram of Drift Memristive Feedforward Layer*

The drift memristor is modelled on FPGA according to formula (9), which $R_{on}$ is set as 14Ω and $R_{off}$ is set as 14kΩ. Based on the model, the circuit diagram of drift memristive feedforward layer is shown as Fig. 10. The input includes clock signals, reset signals, and the pixels of preprocessing images. The preprocessing images, which are represented by currents and range from 0 to 229.5mA, are obtained from the output of the diffusion memristive cellular layers. The controller module determines the state of the entire circuit. The right end of drift memristor model is linked to ground directly during training process. From formula (8), the values of drift memristor models are determined by input current. However, if the values of drift memristor models are tested as ordinary resistors, the values of drift memristor models will be changed, which will affect the accuracy of experimental results. Hence, the right end of drift memristor model is linked to one spike current source whose value is 0.1mA during inference process and the output of the entire circuit is the voltage of drift memristor model. Hence, the values of drift memristor models can be calculated by the output voltage. The entire memristive feedforward layer, which is similar to cellular layer, is pipelined without any complex calculation inside. Hence, the hardware performance of the entire system is good theoretically.

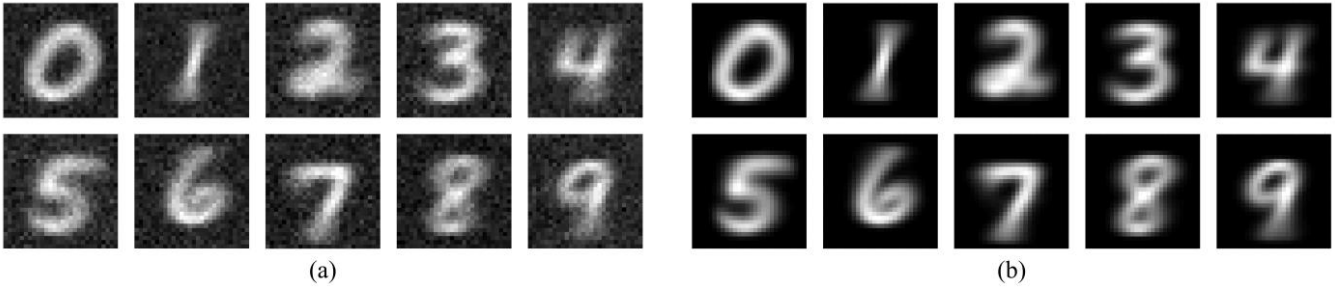

Fig. 13. The images of output current spikes after training (a) without preprocessing (b) with preprocessing

*C. Hardware Optimization and Circuit Diagram of Entire Network*

The circuit diagram of entire network's training and inference is shown as Fig. 11(a) and (b), respectively. The inputs of training process in Fig. 11(a) include clock, reset, and training images (pixels to voltage according to the above). After each training images are processed by diffusion memristive layers and drift memristive layer, the value of drift memristor models won't be changed.

The inference process in Fig. 11(b) use the well-trained drift memristive feedforward layer only. Similar as the training process, the test images are converted into voltage spikes first, but the values of input voltage are a tenth of the input of training process, e. g. input voltage is 1mV for whose pixel value is 100. This spike coding process is to prevent the values of drift memristor models from changing too much by the input of inference process. Put the input voltages (inference images) into the drift memristor models and add the output currents up. Based on Fig. 8 and inference process in Section 3.3, the inference results are got through a comparer, in which the biggest total output currents represents the category of this input inference image.

What's more, for both of training and inference processes, the input are voltage spikes and outputs are current spikes. The proposed multilayer neural network is with in-memory computing and has quite concise and high-efficiency spiking encoding methods.

In order to maximize the processing speed of the entire network, pipeline design methods of hardware including loading and calculation are adopted, which is shown as Fig. 12. Nine pixels are used as a basic unit to perform the training and inference operations. The reason for choosing nine pixels as a basic unit is that it can perform a complete denoising, edge compensation and pattern recognition through layer 1, layer 2, and layer 3, respectively.

IV. EXPERIMENTAL RESULTS AND ANALYSIS

In this section, MNIST dataset with random noise [38] is chosen to test the accuracy of noisy pattern recognition. Further, the entire neural network with image preprocessing and pattern recognition is implemented and optimized on hardware platform to test the performance of hardware resource, processing speed, and throughput.

*A. Pattern Recognition*

Noisy MNIST, which includes 60,000 training images and 10,000 test images with random Gaussian noise $N(0,10^4)$, is chosen as the training and test dataset. The noisy MNIST consists of binary images of 10 numbers from 0 to 9 with 28*28 pixels. To verify the denoising performance, the same voltages is applied to layer 3 to get the output current spikes. The images of output current spikes from layer 3 without preprocessing are shown as Fig. 13(a) and the images with preprocessing are shown as Fig. 13(b). The brighter parts of the output image represent the corresponding pixels which have smaller values of drift memristor models and higher synaptic weights. From above, one can see that the diffusion memristive cellular layer 1 and layer 2 play an important role in denoising of output images. In other words, the image preprocessing of the entire network has a good denoising performance, which lays a solid foundation for pattern recognition task.

TABLE I
IFERENCE RESULTS OF NOISY MNIST

|  |  | Expectation | | | | | | | | | |
|---|---|---|---|---|---|---|---|---|---|---|---|
|  |  | 0 | 1 | 2 | 3 | 4 | 5 | 6 | 7 | 8 | 9 |
| Experiment results | 0 | 968 | 20 | 29 | 23 | 11 | 121 | 11 | 24 | 8 | 19 |
|  | 1 | 0 | 1046 | 1 | 0 | 2 | 5 | 1 | 16 | 0 | 4 |
|  | 2 | 2 | 21 | 919 | 21 | 13 | 6 | 9 | 7 | 2 | 9 |
|  | 3 | 4 | 17 | 19 | 936 | 18 | 23 | 2 | 19 | 3 | 18 |
|  | 4 | 0 | 0 | 7 | 0 | 877 | 0 | 1 | 7 | 2 | 8 |
|  | 5 | 0 | 0 | 0 | 0 | 0 | 719 | 0 | 0 | 0 | 0 |
|  | 6 | 1 | 0 | 11 | 3 | 19 | 0 | 905 | 2 | 6 | 3 |
|  | 7 | 0 | 0 | 2 | 4 | 1 | 1 | 0 | 924 | 1 | 3 |
|  | 8 | 5 | 31 | 44 | 23 | 34 | 17 | 29 | 26 | 950 | 34 |
|  | 9 | 0 | 0 | 0 | 0 | 7 | 0 | 0 | 3 | 2 | 911 |

TABLE II
THE PROCESSING SPEED AND HARDWARE RESOURCE OF ENTIRE NETWORKS

| Device | Processing speed | Hardware resource | Training time | Inference time | Throughput |
|---|---|---|---|---|---|
| Stratix V: 5SGXEA7N2F45C2 | 517.87MHz | 145ALMs | 78.32ms | 1.31μs | 763,358 images/s |

The recognition accuracy of noisy MNIST is shown as TABLE I. The inference accuracy with preprocessing is 91.55% after training. From above, the fully memristive neural network with image preprocessing and pattern recognition is efficient in improving recognition accuracy of noisy MNIST.

*B. Processing Speed and Hardware Resource*

As mentioned above, nine pixels of images are used as a basic unit to perform the training and inference operations. Hence, loading and calculation of one image with 28*28 pixels needs $(28-2)*(28-2) = 676$ times. Specifically, Intel Quartus Prime and Stratix V: 5SGXEA7N2F45C2 are chosen as the software and FPGA platform to test the network's performance, respectively. The processing speed and hardware resource can be obtained from Flow Summary in Quartus Prime after compilation. Through pipeline design, the processing speed and hardware resource are shown as TABLE II. Hence, it takes $(60,000*676)/(517.87*10^6) \approx 78.32$ms to complete the training of 60,000 images and $676/(517.87*10^6) \approx 1.31 \mu s$ to inference one image. The throughput of the entire system is about $1/(1.31*10^{(-6)}) \approx 763,358$ images per second, which is much faster than state-of-the-art works.

*C. Performance Comparison*

To evaluate network's performance on pattern recognition accuracy, hardware resource, processing speed, and throughput, comparison with other works has been done, which is shown as TABLE III. Latest works about memristive neural networks for pattern recognition on hardware are chosen as comparison, including ESIL-MMNN [38], RESPARC [39], CBRAM [40], and HDR-MSN [41]. The criteria include device types, dataset, accuracy, hardware resource, processing speed, and throughput.

ESIL-MMNN [38] demonstrates an in-situ learning with multilayer memristive neural network for pattern recognition. Compared with ESIL-MMNN, our network on noisy MNIST is with roughly the same pattern recognition accuracy as ESIL-MMNN on non-noisy MNIST, benefiting from the image preprocessing function of our network. Further, our network uses less than a third of memristor models in ESIL-MMNN because the multilayer structure in our network needs few devices than the crossbar structure in ESIL-MMNN.

TABLE III
PERFORMANCE COMPARISON WITH STATE-OF-THE-ART WORKS ON MEMRISTIVE SYSTEMS

| Design | Device types | Dataset | Accuracy | Hardware resource | Processing speed | Throughput |
|---|---|---|---|---|---|---|
| ESIL-MMNN [38] | Drift memristors | MNIST | 91.7% | 7,992 memristors | N/A | N/A |
| RESPARC [39] | Drift memristors | MNIST | N/A | 2378 neurons + 1902400 synapses | 200MHz | N/A |
| CBRAM [40] | Drift memristors | MNIST | 92.02% | N/A | N/A | 20 images/s |
| HDR-MSN [41] | Drift memristors | Noisy MNIST | >90% | 3,136 dimensional arrays +784 memristors | <1kHz | <1,000 images/s |
| Our work | Drift and diffusion memristors | Noisy MNIST | 91.55% | 2,352 memristors (145ALMs in Stratix V) | 517.87MHz | 763,358 images/s |

TABLE IV
PERFORMANCE COMPARISON WITH STATE-OF-THE-ART WORKS ON MNIST

| Design | Platform | Resource Usage | Processing speed | Training time | Throughput |
|---|---|---|---|---|---|
| CryptoNets [42] | Intel Xeon E5-1620 CPU | N/A | N/A | N/A | 16.38 images/s |
| LPHS-DL [43] | Titan V GPU | N/A | N/A | 8204.82 (ms) | N/A |
| | Terasic DE1-SoC FPGA | 44856 (LUTs) | N/A | 5091.54 (ms) | N/A |
| HPA-CNN [44] | Xilinx Virtex 7 FPGA | 55774 (LUTs) | 200MHz | 1481.4 (ms) | 40,502 images/s |
| Our work | Stratix V: 5SGXEA7N2F45C2 FPGA | 145 (ALMs) | 517.87MHz | 78.32 (ms) | 763,358 images/s |

RESPARC [39] proposes a memristive crossbar arrays for deep spiking neural networks, which achieves the state-of-the-art processing speed and throughput. However, even the simplest architecture for MNIST in RESPARC still needs 2378 neurons and 1902400 synapses. Compared with RESPARC, multilayer feedforward structure, which is more flexible than crossbar, is applied in our network. Hence, hardware resources of our network are much less than RESPARC. More importantly, our network uses the characteristic of memristor model itself to process the spiking signals without numerous complex calculation modules, so our network achieves higher processing speed and throughput than RESPARC.

CBRAM [40] proposes a digital implementation by low energy subquantum device and achieves high recognition accuracy on MNIST. Compared with CBRAM, the pattern recognition accuracy on noisy MNIST of our network roughly equals with that on pure MNIST of CBRAM, which benefits from image preprocessing function of memristive layer 1 and 2. Further, the throughput of our network is much higher because our network performs a parallel computing system on FPGA rather than using many serial signals for a single image in CBRAM.

HDR-MSN [41] demonstrates a hybrid convolutional neural network based on $HfO_2$ memristor neuron which have a high integration density. Compared with HDR-MSN using the similar noisy MNIST dataset, complex convolutional neural networks are used in HDR-MSN while our network employs spiking signals to implement high-efficiency in-memory computing. Hence, our network has much higher processing speed and over 700 times higher throughput under approximate accuracy and hardware resources.

Furthermore, to show its merits on hardware, some other works [42-44] about MNIST classification on different platforms, including CPU, GPU and FPGA, are chosen as comparisons. CryptoNets [42] focuses on improving the throughput, which is implemented in Intel Xeon E5-1620 and achieves the state-of-the-art performance on CPU. However, CPU has no merits on processing speed of neural networks and deep learning due to its serial computation process. Hence, LPHS-DL [43] on GPU and FPGA, and HPA-CNN [44] on FPGA are chosen as comparisons. Aiming at high processing speed, the LPHS-DL achieves the state-of-the-art training speed on both Titan V GPU and Terasic DE1-SoC FPGA. HPA-CNN is implemented on Virtex 7 FPGA, which is the best 7 series FPGA in Xilinx. Hence, the performance of HPA-CNN, including training time and throughput, is much better than CryptoNets and LPHS-DL.

Compared with the above works on CPU and GPU, the superior performance of our framework benefits from both the learning mechanism and FPGA platform. Compared with HPA-CNN on Xilinx Virtex 7 FPGA, our framework, which is implemented in a similar high-end FPGA (Stratix V: 5SGXEA7N2F45C2), can still achieve better hardware performance including resource usage, processing speed, training time, and throughput.

From above, one can see that our network has a good pattern recognition performance for noisy dataset due to the image preprocessing function of diffusion memristive cellular layers and pattern recognition function of drift memristive feedforward layer. Further, owing to high-efficiency in-memory computing and brief spiking encoding methods, our network needs fewer hardware resources and has much higher processing speed and throughput than state-of-the-art works.

## V. CONCLUSIONS

In this paper, a novel multilayer neural network, which consists of two diffusion memristive cellular layers for image preprocessing and one drift memristive feedforward layer for pattern recognition, is presented. Due to the denoising and edge compensation for preprocessing of diffusion memristive layers, the network has a good anti-noise performance and the recognition accuracy of noisy MNIST is over 90%. Further, because the in-memory computing and spiking encoding methods of the entire network are friendly for hardware implement, the processing speed and throughput are much higher, and the hardware resources are fewer than state-of-the-art works. We hope this idea can give an inspiration for the works combining image preprocessing with machine learning, the applications of diffusion memristor model, and designs of neural networks and neuromorphic computing by integrating memristor models.

## REFERENCES

[1] L. Chua, "Memristor-the missing circuit element," *IEEE Transactions on circuit theory*, vol. 18, no. 5, pp. 507-519, Sep. 1971.

[2] Q. Xia, *et al.*, "Memristor-CMOS hybrid integrated circuits for reconfigurable logic," *Nano Letters*, vol. 9, no. 10, pp. 3640, Sep. 2009.

[3] D. B. Strukov, G. S. Snider, D. R. Stewart, and R. S. Williams, "The missing memristor found," *Nature*, vol. 453, no. 7191, pp. 80-83, May. 2008.

[4] Z. Wang, *et al.*, "Memristors with diffusive dynamics as synaptic emulators for neuromorphic computing," *Nature Materials*, vol. 16, no. 1, pp. 101, Sep. 2016.

[5] Z. Wang, *et al.*, "Resistive switching materials for information processing", *Nature Reviews Materials*, 2020, doi: 10.1038/s41578-019-0159-3.

[6] X. Zhang, *et al.*, "An artificial spiking afferent nerve based on Mott memristors for neurorobotics", *Nature Communications*, vol. 11, art. no. 51, 2020.

[7] B. Linaresbarranco, and T. Serranogotarredona, "Memristance can explain Spike-Time-Dependent--Plasticity in Neural Synapses," *Nature Precedings*, Mar. 2009.

[8] S. H. Jo, *et al.*, "Nanoscale memristor device as synapse in neuromorphic systems," *Nano Letters*, vol. 10, no. 4, pp. 1297-1301, Mar. 2010.

[9] S. Ambrogio, S. Balatti, F. Nardi, S. Facchinetti, and D. Ielmini, "Spike-timing dependent plasticity in a transistor-selected resistive switching memory," *Nanotechnology*, vol. 24, no. 38, pp. 384012, Sep. 2013.

[10] C. Ting, J. Sung-Hyun, and L. Wei, "Short-term memory to long-term memory transition in a nanoscale memristor," *Acs Nano*, vol. 5, no. 9, pp. 7669-76, Aug. 2011.

[11] A. Thomas, "Memristor-based neural networks," *Journal of Physics D: Applied Physics*, vol. 46, no. 9, pp. 093001, Feb. 2013.

[12] S. Kvatinsky, *et al.*, "MAGIC—Memristor-Aided Logic," *IEEE Transactions on Circuits & Systems II: Express Briefs*, vol. 61, no. 11, pp. 895-899, Sep. 2014.

[13] J. J. Yang, D. B. Strukov, and D. R. Stewart,

"Memristive devices for computing," *Nature Nanotechnology*, vol. 8, no. 1, pp. 13-24, Dec. 2012.

[14] X. Zhu, *et al.*, "Ionic modulation and ionic coupling effects in MoS 2 devices for neuromorphic computing", *Nature Materials*, no. 18, pp. 141-148, 2019.

[15] D. Fan, M. Sharad, and K. Roy, "Design and Synthesis of Ultra Low Energy Spin-Memristor Threshold Logic," *IEEE Transactions on Nanotechnology*, vol. 13, no. 3, pp. 574-583, Mar. 2014.

[16] C. Li, *et al.*, "Long short-term memory networks in memristor crossbar arrays", *Nature Machine Intelligence*, no. 1, pp. 49-57, 2019.

[17] Z. Wang, *et al.*, "Reinforcement learning with analogue memristor arrays", *Nature Electronics*, no. 2, pp. 115-124, 2019.

[18] Z. Wang, *et al.*, "In situ training of feed-forward and recurrent convolutional memristor networks", *Nature Machine Intelligence*, no. 1, pp. 434-442, 2019.

[19] E. Covi, *et al.*, "HfO2-based memristors for neuromorphic applications," in *2016 IEEE International Symposium on Circuits and Systems (ISCAS)*, 2016: IEEE, pp. 393-396.

[20] S. Duan, X. Hu, Z. Dong, L. Wang, and P. Mazumder, "Memristor-based cellular nonlinear/neural network: design, analysis, and applications," *IEEE Transactions on Neural Networks & Learning Systems*, vol. 26, no. 6, pp. 1202-1213, Jul. 2015.

[21] J. Moon, *et al.*, "Temporal data classification and forecasting using a memristor-based reservoir computing system", *Nature Electronics*, no. 2, pp. 480-487, 2019.

[22] M. N. Bojnordi and E. Ipek, "Memristive boltzmann machine: A hardware accelerator for combinatorial optimization and deep learning," in *2016 IEEE International Symposium on High Performance Computer Architecture (HPCA)*, 2016: IEEE, pp. 1-13.

[23] F. Cai, *et al.*, "A fully integrated reprogrammable memristor–CMOS system for efficient multiply–accumulate operations", *Nature Electronics*, no. 2, pp. 290-299, 2019.

[24] T. A. Anusudha, and S. R. S. Prabaharan, "A Versatile Window Function for Linear Ion Drift Memristor Model – A New Approach," *AEU - International Journal of Electronics and Communications*, vol. 90, pp. 130-139, Jun. 2018.

[25] Z. Tang, *et al.*, "A hardware friendly unsupervised memristive neural network with weight sharing mechanism," *Neurocomputing*, vol. 332, pp. 193-202, Mar. 2019.

[26] J. H. Yoon, *et al.*, "An artificial nociceptor based on a diffusive memristor," *Nature Communications*, vol. 9, no. 1, pp. 417, Jan. 2018.

[27] Z. Wang, *et al.*, "Fully memristive neural networks for pattern classification with unsupervised learning," *Nature Electronics*, vol. 1, no. 2, pp. 137, Feb. 2018.

[28] M. Hansen, F. Zahari, M. Ziegler, and H. Kohlstedt, "Double-Barrier Memristive Devices for Unsupervised Learning and Pattern Recognition," F*rontiers in Neuroscience*, vol. 11, Feb. 2017.

[29] R. Zhu, *et al.*, "Influence of Compact Memristors' Stability on Machine Learning," *IEEE Access*, no. 7, pp. 47472-47478, Apr. 2019.

[30] R. B. Hur and S. Kvatinsky, "Memristive memory processing unit (MPU) controller for in-memory processing," in *2016 IEEE International Conference on the Science of Electrical Engineering (ICSEE)*, 2016: IEEE, pp. 1-5.

[31] T. Chang, *et al.*, "Synaptic behaviors and modeling of a metal oxide memristive device," *Applied Physics A*, vol. 102, no. 4, pp. 857-863, Mar. 2011.

[32] Z. Tang, *et al.*, "Fully Memristive Spiking-Neuron Learning Framework and Its Applications on Pattern Recognition and Edge Detection," *arXiv preprint arXiv: 1901.05258*, 2019.

[33] Y. V. Pershin, and M. Di Ventra, "Experimental demonstration of associative memory with memristive neural networks," *Neural Networks*, vol. 23, pp. 881-886, Sep. 2010.

[34] C. D. Schuman, *et al.*, "A survey of neuromorphic computing and neural networks in hardware," *arXiv preprint arXiv:1705.06963*, 2017.

[35] K. Cheung, S. R. Schultz, and W. Luk, "NeuroFlow: A General Purpose Spiking Neural Network Simulation Platform using Customizable Processors," *Frontiers in Neuroscience*, vol. 9, no. 19, pp. 516, Jan. 2016.

[36] S. Cawley, "Hardware spiking neural network prototyping and application," *Genetic Programming & Evolvable Machines*, vol. 12, no. 3, pp. 257-280, Sep. 2011.

[37] R. Hu, S. Zhou, Y. Liu, and Z. Tang, "Margin-Based Pareto Ensemble Pruning: An Ensemble Pruning Algorithm That Learns to Search Optimized Ensembles," *Computational Intelligence and Neuroscience*, vol. 2019, art. no. 7560872, Jun. 2019.

[38] C. Li, *et al.*, "Efficient and self-adaptive in-situ learning in multilayer memristor neural networks," *Nature Communications*, vol. 9, no. 1, pp. 2385, Jun. 2018.

[39] A. Ankit, A. Sengupta, P. Panda, K. Roy, "RESPARC:

A Reconfigurable and Energy-Efficient Architecture with Memristive Crossbars for Deep Spiking Neural Networks," in *2017 Proceedings of the 54th Annual Design Automation Conference (DAC)*, 2017: ACM, pp. 1-6.

[40] Y. Shi, *et al.*, "Neuroinspired unsupervised learning and pruning with subquantum CBRAM arrays," *Nature Communications*, vol. 9, Art. no. 5312, Dec. 2018.

[41] J. Wang, *et al.*, "Handwritten-digit recognition by hybrid convolutional neural network based on HfO2 memristive spiking-neuron," *Scientific reports*, vol. 8, no. 1, pp. 12546, Aug. 2018.

[42] N. Dowlin, *et al*., "CryptoNets: Applying Neural Networks to Encrypted Data with High Throughput and Accuracy", in *2016 International Conference on Machine Learning (ICML),* 2016: pp. 201-210.

[43] C. Lammie, M. R. Azghadi, "Stochastic Computing for Low-Power and High-Speed Deep Learning on FPGA", in *2019 IEEE International Symposium on Circuits and Systems (ISCAS),* 2019: IEEE, pp. 1-5.

[44] A. Kyriakos, *et al.,* "High Performance Accelerator for CNN Applications", in *2019 29th International Symposium on Power and Timing Modeling, Optimization and Simulation (PATMOS),* 2019: IEEE, pp. 135-140.